\documentclass[fleqn,twoside]{article}
\usepackage{espcrc2}
\usepackage{url}

\usepackage{graphicx}

\newcommand{\Xmax}{\ensuremath{X_{\rm max}}}
\newcommand{\Ein}{\ensuremath{E_{\rm in}}}
\newcommand{\Eout}{\ensuremath{E_{\rm out}}}

\newcommand{\AmS}{{\protect\the\textfont2
  A\kern-.1667em\lower.5ex\hbox{M}\kern-.125emS}}

\title{Fitting the HiRes Spectra and Monocular Composition }

\author{D.~R.~Bergman
  \address[Rutgers]{Rutgers, 
    The State University of New Jersey  \\
    Department of Physics and Astronomy \\
    Piscataway, New Jersey, USA 08854}
  presented on behalf of the High Resolution 
    Fly's Eye Collaboration}

\begin{document}

\begin{abstract}
  This paper consists of two sections.  In the first section, we
  discuss our fits to the latest HiRes monocular spectra.  We find
  that the best fit for the extragalactic component has a spectral
  index of $\gamma=-2.38\pm0.04$ with a distribution of sources
  varying with a evolution parameter $m=2.8\pm0.3$.  In the second
  section, we discuss preliminary results from a new composition
  measurement using HiRes monocular data.  We find a predominantly
  light spectrum above $10^{17.6}$ eV.  \vspace{1pc}
\end{abstract}

\maketitle

\section{Fitting the HiRes Spectra}

Fitting the ultra high energy cosmic ray (UHECR) spectrum is a common
way to attempt to gain understanding of the sources of these
particles.  These fits often neglect the galactic component of cosmic
rays, which is expected to be small at the very highest energies.
However, the energy at which at which this becomes true is not known.

\begin{figure}[htb]
  \includegraphics[width=\columnwidth]{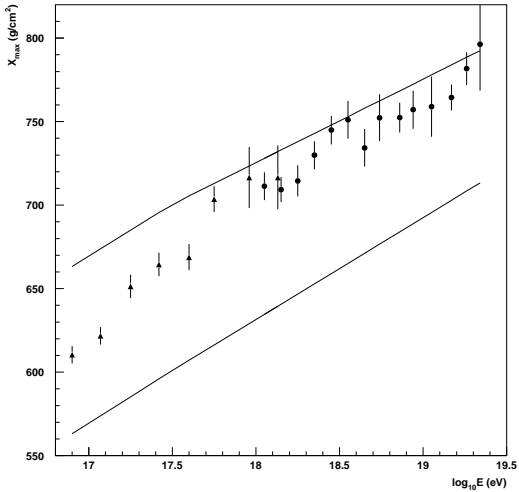}
  \caption{The composition measurements used in the toy model of
    galactic and extragalactic components in the the UHECR spectrum.
    HiRes Prototype/MIA measurements are show as triangles, HiRes
    Stereo measurements as circles.  The lines above and below the
    measurements are the results of
    Corsika/QGSJet\cite{Corsika,QGSJet} calculations for protons and
    iron primaries respectively.}
  \label{fig:comp}
\end{figure}

Rather than assume a negligible galactic component, we employ a toy
model in which composition determines source.  In this model, we use
the recent composition measurements of the HiRes Prototype/MIA
experiment\cite{HiResMIA} and of HiRes in stereo
mode\cite{HiResStereo}.  These are shown in Figure~\ref{fig:comp}.  The
light component in these measurements, assumed to be protons, is
identified with the extragalactic component of UHECRs; the heavy
component, assumed to be iron, is identified with the galactic
component.

Having identified the extragalactic component of the UHECR spectrum,
we fit this component using the assumption that the protons have lost
energy between their acceleration and our observation, and that the
sources are distributed uniformly throughout the universe, with a
density that is smoothly varying in time.  The fits assume a simple
power law spectrum, $E^{-\gamma}$, with a spectral index $\gamma$
identical for all sources.  We use the average energy loss model of
Berezinsky {\it et al.}\cite{Berezinsky} to find the observed spectrum
from a source a given redshift.

In addition, we use a Monte Carlo (MC) simulation\cite{DBO} for the
discrete energy losses associated with pion production.  The MC is
only important for propagating particles from small redshifts,
$z<0.1$, where the proton interacts only a few times.  For propagation
from larger redshifts, the distribution of observed energies for a
given input energy is a gaussian of ever narrowing width.

The MC is run to create an $\Ein$-$\Eout$ matrix with two million
entries for propagation from each redshift, $z$, to our observation.
The $\Ein$-$\Eout$ matrix is calculated every 0.01 step in $z$ for
$z=0$ to $z=4$.  Smaller steps, as small as 0.0001, are used below
0.05.  Sources are assumed to be uniformly distributed at any given
redshift, and with a density with which varies as $(1+z)^m$.  The MC
results at individual redshifts are added together to form an
$\Ein$-$\Eout$ matrix for uniformly distributed sources evolving
according given value of $m$.  These $z$-integrated matrices were
formed for $m=$0 to $m=$5 in steps of 0.5.

The fits are performed by varying the normalization, the spectral
slope ($\gamma$), and the evolution parameter ($m$).  The
goodness-of-fit is evaluated using the binned-maximum-likelihood
method.  The comparison spectra for fitting are calculated by
multiplying a vector containing the input spectrum by the
$\Ein$-$\Eout$ matrix using matrix multiplication.  The input spectrum
vectors are calculated every 0.05 in the spectral index $\gamma$.  The
comparison spectra for intermediate values of $\gamma$ and $m$ are
found by interpolation.

\begin{figure}[htb]
  \includegraphics[width=\columnwidth]{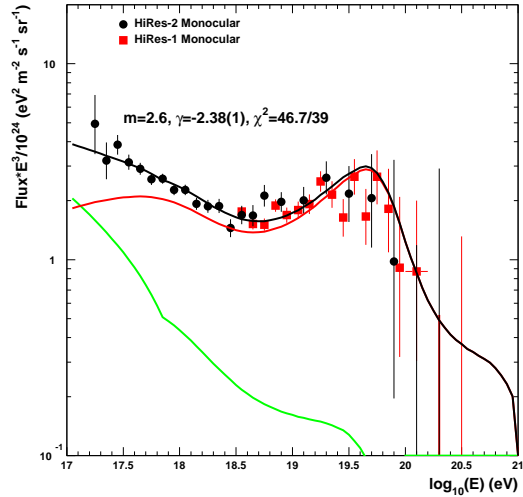}
  \caption{The HiRes-I and HiRes-II monocular spectra, with the result
    of the best fit spectrum.  HiRes-I points are shown as red
    squares, HiRes-II points as black circles.  The one $\sigma$ upper
    limit on the flux in two more bins for each measurement are shown
    above the highest energy actually observed.  The black fit line is
    the sum of the galactic (green) and extragalactic (red)
    components.  The parameters of the extragalactic component are
    $\gamma=-2.38$ and $m=2.6$.  There are 42 data points, giving 39
    degrees of freedom in the fit.}
  \label{fig:bestfit}
\end{figure}

\begin{figure}[htb]
  \includegraphics[width=\columnwidth]{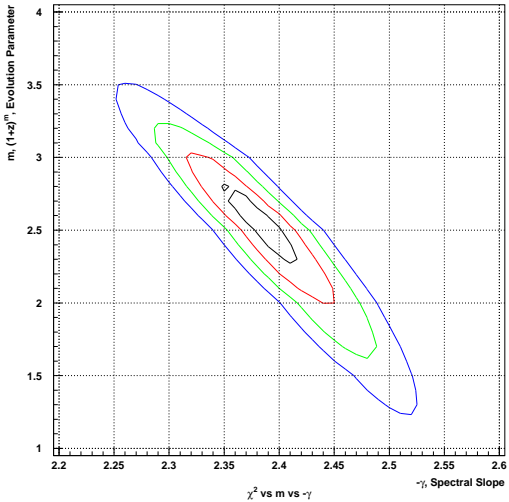}
  \caption{A scan of $\chi^2$ ($-2\log L$) in $m$-$\gamma$ space.  The
    first four $\sigma$ contours are shown.}
  \label{fig:ch2plot}
\end{figure}

The result of the fit to the most recent HiRes-I\cite{HR1} and
HiRes-II\cite{HR2} monocular spectra is $\gamma=-2.38\pm0.04$ and
$m=2.8\pm0.3$.  This fit is shown in Figure~\ref{fig:bestfit}.  The
quoted uncertainties include the effect of the correlation between
$\gamma$ and $m$.  This correlation is shown in
Figure~\ref{fig:ch2plot}.  Not included in this fit are any systematic
uncertainties.  The largest systematic uncertainty is expected to be
due to the details of composition used in the aperture calculation for
the HiRes-II spectrum.  Changes in the composition tend to move the
best fit point along the trough of Figure~\ref{fig:ch2plot}, changing
the $m$ vs.  $\gamma$ relationship little.

\section{HiRes Monocular Composition Measurement}

Because the uncertainty in composition is a large part of the
uncertainty in the parameters in the fit to the spectrum, we were
motivated to try to improve on the existing composition measurements,
especially below $10^{18}$ eV.  To go to low energies with HiRes
requires using the monocular data sets, and we chose the HiRes-II
monocular data set for this analysis.

Unfortunately, the limited elevation coverage of the HiRes-II aperture
biases our $\Xmax$ acceptance, with the bias increasing at lower
energies.  The bias stems from the requirement that we find $\Xmax$
within that extent of the shower observed in the detector.  Events
that are closer to the detector are more likely to have $\Xmax$ above
the visible range, and thus be cut.  Furthermore, because lower
energies can only be observed close to the detector, events at these
energies will have a larger acceptance bias than those at higher
energies.  This bias precludes performing an elongation rate analysis
at energies below $10^{18}$ eV.

\begin{figure}[htb]
  \includegraphics[width=\columnwidth]{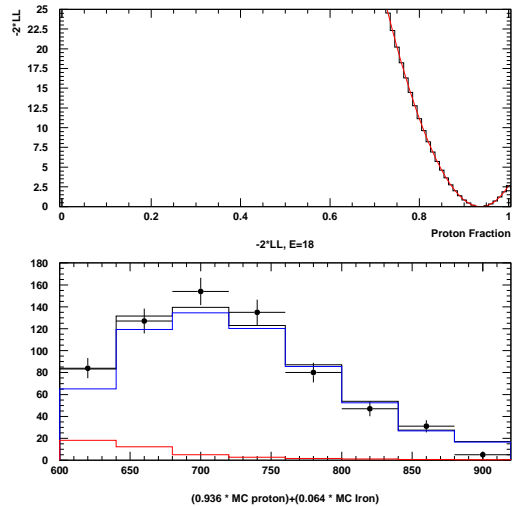}
  \caption{The results of a fit to \Xmax\ at an energy of $10^{18}$
    eV.  The top panel shows $-2\log L$ as a function of the proton
    fraction in 1\% steps (black histogram), and a polynomial fit to
    the same (red line).  The bottom panel shows the \Xmax\ 
    distribution that results from the fit, with the data shown as
    filled circles with error bars, the proton distribution in blue,
    the iron distribution in red and the sum in black}
  \label{fig:compfit18}
\end{figure}

\begin{figure}[htb]
  \includegraphics[width=\columnwidth]{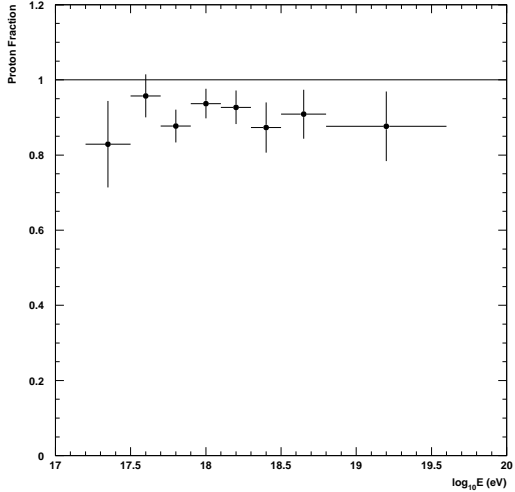}
  \caption{The raw composition measured in the data.}
  \label{fig:rawdata}
\end{figure}

Instead, we have chosen to fit the \Xmax\ distribution in energy bins
to a combination of MC generated \Xmax\ distributions from proton
primaries and from iron primaries.  The \Xmax\ distribution in each
energy bin is stored in a histogram, and the data histogram is then
compared to a mixture of the proton and iron MC histograms to find the
best fit proportion of each.  The fit is performed using the binned
maximum likelihood technique as implemented in the HBOOK\cite{hbook}
routine HMCLNL.  The uncertainty of the fit in each energy bin is
taken from the width of $-2\log L$ as fit to a quadratic in the region
about the maximum.  An example of this procedure for the bin at
$10^{18}$ eV is shown in Figure~\ref{fig:compfit18}.  The result of
these fits for the data at all energies is shown in
Figure~\ref{fig:rawdata}.

\begin{figure}[htb]
  \includegraphics[width=\columnwidth]{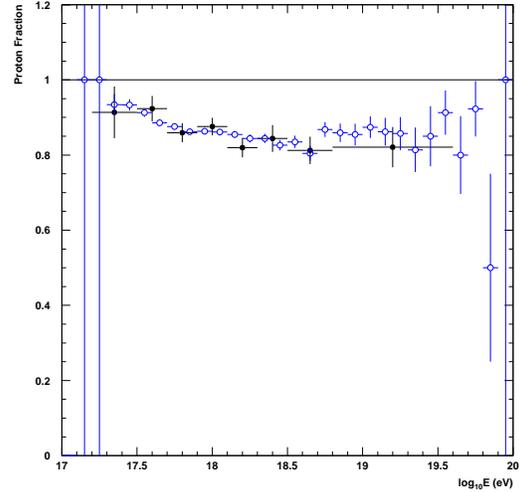}
  \caption{The raw composition measured in the MC sample used in the
    spectrum calculation.  The black, filled circles are the result of
    the fits.  The blue, open circles are the actual fractions using
    the known,  {\it a priori} information available in the MC sample.}
  \label{fig:rawmc}
\end{figure}

This fit gives only the proportion of proton and iron in the final,
accepted sample; as such, it is not intrinsically interesting, but
must be corrected to find the proton-iron ratio of UHECRs entering the
atmosphere.  However, one can test this procedure on an independent MC
sample where one knows {\it a priori} which events are protons and
which are iron.  The result of this analysis on the MC sample used to
calculate the aperture used in the spectrum analysis is shown in
Figure~\ref{fig:rawmc}.  The fits agree very well with what is
actually in the sample.

The raw composition measurements are corrected using the relative
acceptances in the proton and iron MC samples.  These are the same MC
samples used to provide the comparison \Xmax\ distributions used in
the fit.  The fit proton-iron proportion is interpreted as given
numbers of proton events and iron events.  These numbers then are
separately corrected for the relative acceptance appropriate to each.
Finally, the proton-iron ratio is recalculated using the corrected
numbers of events of each type.

\begin{figure}[htb]
  \includegraphics[width=\columnwidth]{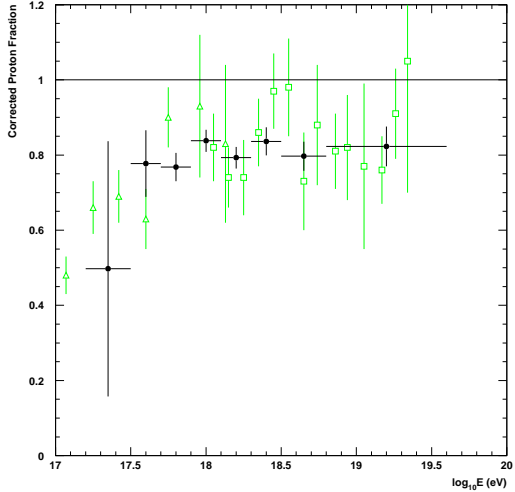}
  \caption{The corrected composition measured in the MC sample used in the
    spectrum calculation.  The black, filled circles are the corrected
    result of the fits.  The green, open triangles and squares are the
    HiRes Prototype/MIA and HiRes Stereo composition measurements
    which form the input to the MC.}
  \label{fig:cormc}
\end{figure}

For a MC sample, we can again compare the corrected proton-iron
proportion to the input to the MC simulation.  This is shown in
Figure~\ref{fig:cormc}.  In this case, the input is just the HiRes
Prototype/MIA and HiRes Stereo composition measurements, where the mean
\Xmax\ of the data is interpreted as representing a proton-iron
proportion given by the distance between the Corsika/QGSJET proton and
iron lines.  The input is in fact given by linear fits to the mean
\Xmax\ above and below $10^{17.85}$ eV.  Again, the corrected results
of the fits agree very well with the inputs to the MC.

\begin{figure}[htb]
  \includegraphics[width=\columnwidth]{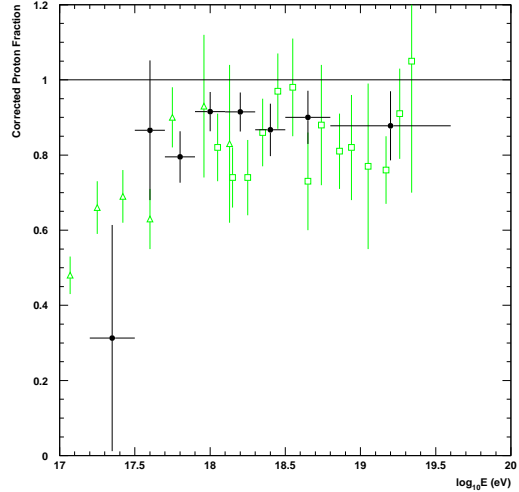}
  \caption{The preliminary, corrected composition measured in the
    data.  The black, filled circles are the corrected result of the
    fits.  For comparison, the HiRes Prototype/MIA and HiRes Stereo
    composition measurements are shown as green, open triangles and
    squares, respectively.}
  \label{fig:cordata}
\end{figure}

Finally, we show as a very preliminary result, the corrected
proton-iron ratio for the data.  See Figure~\ref{fig:cordata}.  This
result shows a very light composition (90\% protons) above $10^{17.6}$
eV.  Below this energy we measure a sharply lower composition, but
with large uncertainties because of few events and a large acceptance
correction.  The measured composition at high energies is in agreement
with the HiRes Stereo measurement, but closer to being all protons.

The data used in the fits comes from the same data set used in the
HiRes-II spectrum analysis which was fit above.  A considerable amount
of data, about a factor of three more, was collected after this data
set.  This later data was collected with lower light thresholds, and
as such, will have a lower energy threshold.  In addition, the
aperture at each energy will be increased, reducing the size of \Xmax\ 
biases.  Performing this analysis on this expanded data set should
give us much tighter constraints on the composition in the middle of
the $10^{17}$ eV energy decade.

The Telescope Array (TA), which has already received Japanese funding,
and the TA Low Energy extension (TALE) will be able to push this
analysis even further.  TALE will use the HiRes phototubes with larger
mirrors to push the thresholds even lower.  The mirrors will be
arranged in a tower rather than a ring to avoid biases in \Xmax.
Geometrical resolution at the level of HiRes Stereo will be provided
by coincidences with the TA ground array.

\end{document}